\documentclass[12pt,psfig]{article}
\usepackage{amssymb}
\usepackage{amsmath}
\usepackage{a4wide}
\usepackage{epsfig}
\def\be{\begin{equation}}
\def\ee{\end{equation}}
\def\bea{\begin{eqnarray}}
\def\eea{\end{eqnarray}}
\newcommand{\f}{\frac}

\begin{document}
\begin{titlepage}
\title{{\bf An operational view on the holographic information bound}} \vskip2in

\author{
{\bf Karel Van Acoleyen\footnote{\baselineskip=16pt E-mail: {\tt
karel.vanacoleyen@ugent.be }} }
\hspace{3cm}\\
 $$ {\small Department of Physics and Astronomy, Ghent
 University,}\\
 {\small Krijgslaan 281, S9, 9000 Gent, Belgium}.
}

\date{}
\maketitle
\def\baselinestretch{1.15}
\begin{abstract}
\noindent We study the covariant holographic entropy bound from an
operational standpoint. Therefore we consider the physical limit
for observations on a {\em light-sheet}.  A light-sheet is a
particular null hypersurface, and the natural measuring apparatus
is a screen. By considering the physical properties of the screen
- as dictated by quantum mechanics - we derive an uncertainty
relation. This connects the number of bits of decoded information
on the light-sheet to two geometric uncertainties: the uncertainty
on the place where the bits are located and the uncertainty on the
local expansion of the light-sheet. From this relation we can
argue a local operational version of the (generalized) covariant
entropy bound: the maximum number of bits decoded on a light-sheet
interval goes like the area difference (in Planck units) of the
initial and final surface spanned by the light rays.

\end{abstract}

\thispagestyle{empty} \vspace{5cm}  \leftline{}

\vskip-19cm  \vskip3in

\end{titlepage}
\setcounter{footnote}{0} \setcounter{page}{1}
\newpage
\baselineskip=20pt

\section{Introduction}

The holographic information bound states that, at the fundamental
level, the information content of a volume is bounded by the area
of its boundary surface at one bit per Planck area
\cite{'tHooft:1993gx}. This scaling of information with the
surface area rather than the volume, is clearly at odds with any
expectation from local quantum field theory. Nevertheless it seems
to be a fact of nature. And as such, it appears to reveal an
important, yet mysterious aspect of the true nature of (quantum)
gravity.

Actually, proclaiming the holographic information bound as a fact
of nature, requires a carefull formulation. This formulation was
given by Bousso and goes by the name of the covariant entropy
bound \cite{Bousso:1999xy}. It is this bound that has been
empirically validated for a variety of physical systems, ranging
from black holes and collapsing stars, to entire FLRW universes
\cite{Bousso:2002ju,Bousso:2010pm}. To state the covariant entropy
bound, one can pick an arbitrary two-dimensional spatial surface.
This surface then lies on the boundary of a {\em light-sheet} $L$.
This is the null hypersurface, that is generated by
surface-orthogonal null geodesics, in the contracting
(non-expanding) direction from the boundary surface. A light-sheet
(locally) terminates when a caustic or singularity is reached. The covariant
entropy bound now states for the entropy $S$ on $L$:\footnote{We
work in natural units $c=\hbar=k_b=1$, but we will keep track of
the Planck length $l_P=\sqrt{G_N}$ in our equations.} \be S(L)\leq
\f{A}{4l_p^2}\,.\label{CEB}\ee

So far, no general derivation of this bound has been given.
However, Flanagan, Marolf and Wald were able to formulate two sets
of general assumptions about the relation between entropy density
and energy density, from which the bound can be derived
\cite{Flanagan:1999jp}. Furthermore, they showed that one set of
their assumptions implied an even stronger bound, for the entropy
on general {\em light-sheet intervals}. Consider a light-sheet
that starts off from a surface $B$ with area $A$. Now, instead of
following the light rays\footnote{In this paper, we use the term
'light ray' as a synonym for 'null-geodesic', not to denote an
actual photon path.} all the way to the caustics, we can terminate
the light-sheet sooner, with the rays spanning a surface $B'$,
with non-zero area $A'<A$. The generalized covariant entropy bound
(GCEB) then states for the entropy on this light-sheet interval:
\be S\leq \f{A-A'}{4 l_P^2}\label{SCEB}\,. \ee Clearly, the
covariant bound (\ref{CEB}) is a special case of (\ref{SCEB}). And
also the stronger bound seems to hold for physical systems.
There were some presupposed counterexamples for short light-sheet
intervals \cite{Bousso:2002ju}. But it was shown that these
examples signalled a failure of the local description of entropy
at short distances, rather than a failure of the GCEB
\cite{Bousso:2003kb}.

In this paper we derive an operational version of this stronger
entropy bound. For this we need a suitable interpretation of the
entropy in (\ref{SCEB}).  As was stressed by Bousso, the covariant
entropy bound does not single out a preferred time direction
\cite{Bousso:1999xy}. Therefore the entropy in (\ref{SCEB}) has to
be truly statistical in nature, referring to the amount of
microscopic information \cite{Jaynes:1957zza}. We can therefore
interpret $S=\Delta N$, as the number of (nat)bits that specify a
particular light-sheet interval. From the operational standpoint
this becomes the number of bits that can be acquired by physical
measurements on the particular light-sheet interval. From some
simple arguments that solely depend on elementary quantum
mechanics and (classical) gravity we will argue that this number
does indeed obey the bound (\ref{SCEB}). As measuring apparatus we
will consider a screen. We will later substantiate to some extent
that a screen does indeed represent an optimal measuring device
for light-sheets.

\section{The screen} To capture the balance between energy cost and information
capacity that is set by quantum mechanics, we will model an
idealized screen. This follows largely the construction of
\cite{Dvali:2008ec} (section 3).  We consider the screen
consisting of $N_{pix}$ individual pixels with size $l_{pix}$.
This pixel size, sets both the space and time resolution of the
screen. We will sometimes also use $\Lambda=1/l_{pix}$ and we can
think of this as the energy resolution or the bandwidth of the
screen. The screen has a surface area $A=N_{pix}l_{pix}^2$, but it
will be vital to recognize that it has a thickness as well, of the
order $l_{pix}$. Furthermore, as a check on the generality of some
of our results, we explicitly take into account the number of
species $N_s$ that can encode information. This does not matter
much for a screen that is only sensitive to photons for instance,
but it will be important when the number of species gets large,
like in the case of the 'CFT screen' on AdS \cite{work in
progress}. The maximum number of bits that can be decoded on the
screen in a time interval $l_{pix}$ will be: \be \Delta
N=N_sN_{pix}=N_s\f{A}{l_{pix}^2}\,.\label{deltabits} \ee As most
of our equations from now on, this has to be interpreted as an
estimate, for which the precise (order one) pre-factor is
undetermined.

For each pixel, the screen needs the ability to recognize each of
the $N_s$ species. This can be achieved for instance, by storing
$N_s$ sample particles in every pixel, that can interact with -
and therefore detect - every type of particle. The minimal energy
of a sample particle localized within a pixel of size $l_{pix}$,
is $E=1/l_{pix}$, so every pixel will necessarily store an energy
$E= N_s/l_{pix}=N_s\Lambda$. Conversely, the energy of a sample
particle sets a minimal thickness of the pixel - and therefore the
screen - of the order $l_{pix}$, as we mentioned above. Furthermore, neglecting boundary
effects, we can write down an expression for the (coarse-grained)
energy-momentum tensor of the pixel: \be T_{\mu\nu} = N_s
\Lambda^4 (U_{\mu}U_{\nu}- k g_{\mu\nu})\,,\label{emt}\ee where
$U_{\mu}$ is the four-velocity of the pixel (screen) and with $k$
an order one constant.

Finally, as was argued in \cite{Dvali:2008ec}, the required energy
for the pixel sets an intrinsic maximum resolution. Indeed, when
the energy of the pixel becomes too large, it collapses to a black
hole. This results in the bound:   \be l_{pix}^2 \geq
l_{N_s}^2\equiv N_s l_P^2\,.\label{intr}\ee For a few species,
this is simply the operational derivation of $l_P$ as the smallest
length scale; for a large number of species this is the hierarchy
between the smallest length scale $l_{N_s}$ at which individual
species can be resolved and the Planck length.

\section{Observing on the light-sheet}
We set things up, so that we can use our screen to make a
measurement on a particular light-sheet. This can be done in an
obvious way, by positioning and orienting the screen such that the
light rays of the light-sheet at some value $\lambda$ of the
affine parameter, will be orthogonal to the screen at
some time $t$ in the screen reference frame. At this instant, the
screen coincides with part of the surface $B(\lambda)$ spanned by
the light rays of the sheet. Notice that we are assuming that the
size of our screen is smaller than the smallest distance scale set
by the intrinsic and extrinsic curvature of the surface on the
light-sheet, so that we can work in the local flat limit. Since a
screen may consist of only a few pixels, this is in fact a
condition on the pixel size $l_{pix}$. Now, if the measurement
would be really instantaneous and the screen would be truly a
two-dimensional object, we could assign all the information acquired
during the measurement to (part of) the surface $B(\lambda)$ of
the specific light-sheet. But clearly this is not the case. As we
discussed above, the screen necessarily has a non-zero space-time
resolution $l_{pix}$, and a thickness of the same order. As
illustrated in the figure, this carries through to a smallest
interval $\Delta \lambda$ of the affine parameter, over which we
can locate the measured information on the {\em coarse-grained}
light-sheet. We will not further specify the coarse-graining procedure, and leave it at the simple picture of the figure. In fact, in what follows we will drop the adjective and simply talk of the light-sheet. With $k^{\mu}=dx^\mu\!/d\lambda$, the
local tangent vector to the light rays (and $U^\mu$ the
four-velocity of the screen), the covariant expression for the
light-sheet interval reads: \be \Delta \lambda
 = \f{l_{pix}}{|k_{\mu}U^\mu|}\,.\label{deltalam}\ee This holds at first order in $\Delta \lambda$, as will be the case for the other derivations that follow.

\begin{figure}
\begin{center}
\includegraphics[width=7cm]{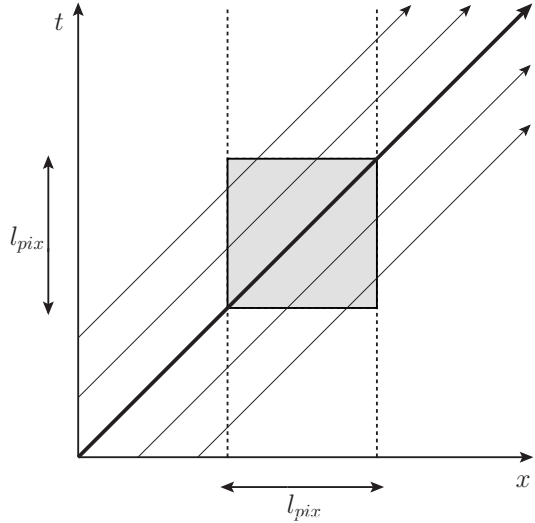}
\caption{A measurement in the screen rest frame, with the
$x-$direction orthogonal to the screen. The
measurement takes place in the shaded space-time region set by the
pixel size. Correspondingly, we will consider the acquired information to be located on the coarse-grained light-sheet (depicted as a thick light ray), over an interval
$\Delta \lambda$. }
\end{center}
\end{figure}

Since we are interested in the maximum number of bits that can be
decoded, we will estimate the maximum possible resolution of our
screen. Of course, we first of all have the intrinsic bound
(\ref{intr}), set by the screen itself. But, as we will now argue,
there is another bound, that is set by the particular light-sheet
that we want to observe. It is the latter bound that will
translate to an operational version of the GCEB.

Our heuristic argument on this bound follows the mantra of the
classic quantum mechanics gedanken experiments on the uncertainty
relations. Namely, that one should always consider the influence
of the measuring apparatus (in our case the screen) on the object
being measured (in our case the light-sheet). Clearly, the screen
has energy, energy bends light rays, therefore the light-sheet is
disturbed by the screen. To quantify this effect we should look at
the expansion of the family of light rays that generate the
light-sheet. The expansion $\theta$ of a family of light rays with
tangent vector $k^{\mu}$ is defined as (see e.g.
\cite{Bousso:2002ju,Wald}): \be \theta(\lambda)=\nabla_\mu
k^{\mu}=\f{1}{\mathcal{A}}\f{d\mathcal{A}}{ d\lambda}\,,\ee where
$\mathcal{A}$ is the surface area spanned by the infinitesimally
neighboring light rays. Its evolution is governed by the (zero
twist) Raychaudhuri equation: \be
\f{d\theta}{d\lambda}=-\f{1}{2}\theta^2-\sigma_{\mu\nu}\sigma^{\mu\nu}
-8\pi G_N T_{\mu\nu}k^{\mu}k^{\nu}\,.\label{Raychaudhuri}\ee The
second term on the right-hand side is always non-positive, and
gives the focussing contribution of the shear. The third term is
non-positive if matter obeys the {\em null energy condition},
$T_{\mu\nu}k^\mu k^\nu\geq 0$. This term then captures the
focussing effect of the total energy-momentum and the screen
influences the expansion in the first place through its
contribution to this term.\footnote{The screen also influences the evolution of the shear through its contribution to the Weyl curvature. But one can show that the resulting local effect on $\Delta \theta$ is only of order $\Delta \lambda ^2$.} The screen contribution is nonzero only
in the interval $\Delta \lambda$, and we can estimate the
resulting 'kick' $\Delta \theta$ of the expansion by plugging in
the expression (\ref{emt}) for the energy-momentum tensor of the
screen and using our expression (\ref{deltalam}) for
$\Delta\lambda$: \be \Delta\theta=-N_s l_P^2\Lambda^4 (k_\mu
U^\mu)^2 \Delta \lambda = -\f{N_s l_P^2 |k_\mu
U^\mu|}{l_{pix}^3}\,. \label{deltatheta} \ee We can now interpret
$-\Delta \theta$ as the uncertainty on the expansion of the
light-sheet, associated with the screen measurement.
Inspection of the full equation (\ref{Raychaudhuri}) shows that,
at best, this estimate holds for $\Delta \lambda\leq 1/|\theta|$,
with $\theta$ the undisturbed expansion. Notice that this range
also corresponds to the maximal possible range on the light-sheet,
since by the {\em focussing theorem} we have that
$\,\theta(\lambda+\Delta\lambda)\rightarrow -\infty\,$ for
$\,\Delta \lambda \leq 2/|\theta(\lambda)|\, $, as one can easily show from
(\ref{Raychaudhuri}) (see e.g. \cite{Bousso:2002ju}).

So we find three relevant quantities that depend on the screen resolution, velocity and number of decoding species. There are the uncertainties on $\theta$
(\ref{deltatheta}) and on the value of the affine parameter
$\lambda$ (\ref{deltalam}). And there is the maximum number of
bits $\Delta N$ (\ref{deltabits}) that can be acquired in the
measurement. Taken together, we can write this interdependence in
the following uncertainty relation (dropping the minus sign in
(\ref{deltatheta})): \be \Delta\theta\,.\,\Delta\lambda\geq
\f{l_P^2}{A}\Delta N\,, \label{uncertainty}\ee
 the uncertainty on the local expansion multiplied by the uncertainty on the affine parameter of the light-sheet is larger than the number of bits decoded on the corresponding light-sheet interval, per spanned surface area in Planck units. Note that this relation is covariant and does not depend on the particular affine parameterization. Nor does it depend on the different screen properties, which suggests a universal character.

It is now a small, albeit heuristic step from the uncertainty
relation above, to an operational version of the GCEB
(\ref{SCEB}). In the same way that one argues from Heisenberg's
uncertainty principle, that $1/m$ is the smallest wavelength that
can be used to resolve a particle with mass $m$; we can now argue
the largest resolution allowed for the screen to observe at a
particular place of the light-sheet. We simply require the
measurement kick or uncertainty on the local expansion, to be less
than the expansion itself. This leads to the condition: \be \Delta
\theta = |\theta| \,,\label{condtheta}\ee for the maximum
resolution of the screen observing on a light-sheet with expansion
$\theta$ at the cross-section with the screen. For observations at
this maximum resolution, the uncertainty relation
(\ref{uncertainty}) becomes:
  \be \Delta N \leq\f{A}{l_P^2}\, |\theta|\,\Delta\lambda=\f{1}{l_P^2}\Delta A \,, \label{opSCEB}\ee
the maximum number of bits that can be decoded on a light-sheet
interval is smaller than the change in the surface area spanned by the light rays, in Planck units. Apart from the constant $1/4$, that can not be determined from our argument, we consider this to be the {\em local} operational version of the GCEB (\ref{SCEB}). Local in the sense that our derivation holds up to first order in $\Delta \lambda$. As such, our operational construction is also too rudimentary to go beyond this order.

For the validity of the covariant entropy bound one usually
assumes two energy conditions on the matter sector
\cite{Bousso:2002ju}. First, there is the null energy condition
that we quoted before. Secondly, one assumes the {\em causal
energy condition} which excludes superluminal propagation of
energy. In our derivation of the operational bound we have
implicitly used the very same two energy conditions. Indeed, when
modelling the screen, we assumed that there were no {\em ghosts},
i.e. quanta with negative energy, that are excluded by the null energy condition. Those particles would make it
possible to store information without the usual energy cost.
In addition, when converting the spatial resolution to a time resolution we use
the speed of light. This would obviously change in the case of
superluminal propagation.

Furthermore, the covariant entropy bound is considered to be a
classical bound in the sense that it only applies to those regimes
where the space-time geometry can be treated as approximately
classical. This is also the case for our operational bound. But
notice that the bound itself follows precisely from recognizing
that the concept of absolute classical space-time geometry does
not make sense from an operational point of view. Indeed,
according to our uncertainty relation (\ref{uncertainty}), one can
not both determine the local expansion and the corresponding
affine parameter up to an arbitrary accuracy.

The physical mechanism behind the operational bound is clear. A
measurement that allocates $\Delta N$ bits of information to a
certain place on the light-sheet, necessarily comes with an
uncertainty $\Delta \lambda$ on the precise location of these
bits, such that the corresponding light-sheet interval obeys the
bound (\ref{opSCEB}). To see in detail how this comes about for
measurements with our screen, we have to look at the maximum
screen resolution, at which the bound (\ref{opSCEB}) can be
saturated. We will call this {\em the holographic resolution}.
From the condition (\ref{condtheta}) and the expression
(\ref{deltatheta}) for $\Delta \theta$, we find \be
l_{pix}^3=l_{hol}^3\equiv \f{N_s
l_P^2|k_{\mu}U^{\mu}|}{|\theta|}\,,\label{holres}\ee at this
maximum resolution, for a screen with $N_s$ species and
four-velocity $U^\mu$, observing on a light-sheet locally
generated by $k^\mu$, with expansion $\theta$.  As before, this
formula is covariant and independent of the particular affine
parameterization. But notice that it does depend on the screen
velocity. By varying the speed of the screen in the direction of
the light-rays from $c$ to $-c$, the factor $|k_\mu U^\mu|$ varies
from zero to infinity. Suppose now we have a screen with
resolution $l_{pix}$ and we want this screen to operate at the
holographic resolution for a given light-sheet with local
expansion $\theta$. This requires a screen velocity such that: \be
|k_\mu U^\mu|=\f{|\theta|}{N_s l_P^2}l_{pix}^3\,,\ee and for the
resolution $\Delta\lambda$ of the light-sheet position we then
find: \be \Delta \lambda=\f{l_{pix}}{|k_\mu U^\mu|}=\f{N_s
l_P^2}{|\theta|\,l_{pix}^2}\,.\label{deltalam2}\ee The maximum
number of decoded bits $\Delta N$ (\ref{deltabits}) then indeed
obeys the operational bound (\ref{opSCEB}). It is now interesting
to look what happens for a screen that operates at maximum
intrinsic resolution $l_{pix}^2=N_s l_P^2$. As one sees from
(\ref{deltabits}), such a screen can acquire one bit per Planck
area in a measurement. This corresponds to Bousso's bound for
entire light-sheets. Our estimate (\ref{deltalam2}) of the
associated light-sheet interval then reads $\Delta \lambda
=1/|\theta|$. Which, as we discussed earlier, corresponds to the
maximum possible range for which our derivation holds and which is
also simply the maximum possible range on the light-sheet.

Clearly, the most heuristic step in our derivation of the
operational information bound is the condition (\ref{condtheta}),
that determines the holographic resolution. Let us briefly
illustrate its physical meaning for the simplest example. We
consider (static) spherical screens that observe on the
light-sheets emanating from some spherical surface in Minkowski
space-time. Substituting the expansion and velocity factor in
(\ref{holres}), we find the holographic resolution $l_{pix}^3=N_s
l_P^2 r$ for a screen at a distance $r$ from the origin. The
screen has a thickness $l_{pix}$ and therefore a volume $r^2
l_{pix}$. Using the expression (\ref{emt}) for the density,
$\rho=N_s/l_{pix}^4$, this leads to a mass $M= r/l_P^2$. So in
this case, our condition (\ref{condtheta}) simply states that the
screen itself should not be a black hole. In \cite{work in
progress} we will provide further illustrations on other
space-times.

Of course, our derivation of the general uncertainty relation
(\ref{uncertainty}) and the corresponding operational bound
(\ref{opSCEB}), also crucially hinges on the assumption that the screen
is an optimal measuring apparatus for light-sheets. We can not
prove this. However, we can consider a slight variation of our
screen and show that it performs worse, in the sense that it can
not saturate the bound (\ref{opSCEB}). Let us take a multi-layered
screen consisting of a stack of $n$ ordinary screens. Both the
kick in the expansion $\Delta \theta$, as the interval of the
affine parameter $\Delta \lambda$ (but not the resolution), as the
number of bits $\Delta N$ per measurement, scales linearly with
$n$. Following the same  logic that we applied before, this leads
to an extra factor $1/n$ on the right-hand side of the inequality
(\ref{opSCEB}), for observations with this multi-layered screen.

Finally, the generalization of our results to arbitrary $d+1$
space-time dimension is straightforward. The uncertainty relation
(\ref{uncertainty}) and the bound (\ref{opSCEB}) for example,
remain of the same form, with on the right-hand side,
$A/l_P^{d-1}$ denoting the surface area of the relevant
codimension 2 surface in Planck units. While the formula
(\ref{holres}) for the holographic resolution becomes: \be
l_{hol}^{d}= \f{N_s
l_P^{d-1}|k_{\mu}U^{\mu}|}{|\theta|}\,.\label{holresgen}\ee

\section{To Conclude}
In this paper we have in some sense provided an approximate {\em
derivation} of the (generalized) covariant entropy bound. But even if
we could go beyond our local approximation, and even if we could
determine the precise pre-factors, we would not consider this to
constitute an {\em explanation}. The holographic entropy bound is
such a beautiful relation that it would be surprising if it's
merely a consequence of quantum mechanics and classical gravity.
One would expect there to be new physics. The case in point being
the AdS/CFT connection. Maybe our operational approach will help
in unveiling some of the new physics for other, more general
space-times. In this light we can not resist to point out the
analogy with the relation between Bekenstein's entropy bound
\cite{Bekenstein:1980jp} and Heisenberg's uncertainty relation.
In that case one can interpret the bound as the information
theoretic translation of the uncertainty relation
\cite{Bousso:2004kp}. In the same way one could view the generalized covariant
entropy bound as the translation of our uncertainty relation
(\ref{uncertainty}).

\section{Acknowledgments}
It's a pleasure to thank Frank Verstraete and especially Henri Verschelde for stimulating and encouraging discussions. Thanks as well to Eanna Flanagan, Ignacio Navarro and Joachim De Beule, for their comments on a draft version of the paper. I am supported by a postdoctoral grant of the Fund for Scientific Research Flanders (Belgium).

\end{document}